\documentclass[a4, 11pt]{article}

\usepackage{amsmath, amssymb, amsthm, bm, graphicx, color, lineno, setspace}
\usepackage[comma]{natbib}

\newcommand{\fulltoday}{\number\day\space \ifcase\month\or
  January,\or February,\or March,\or April,\or May,\or June,\or
  July,\or August,\or September,\or October,\or November,\or December,\fi
  \space\number\year}

\newcommand{\E}[2][]{{\rm E}_{#1}\left[#2 \right]}

\newcommand{\Cov}[1]{{\rm Cov}\left[#1 \right]}

\theoremstyle{definition}

\title{Measuring temporal turnover in ecological communities\footnote{This is authors' version pre-print of: 
Hideyasu Shimadzu, Maria Dornelas and Anne E. Magurran (2015) Measuring temporal turnover in ecological communities. {\it Methods in Ecology and Evolusion} {\bf  6}(12): 1384--1394. \mbox{doi: \texttt{http://dx.doi.org/10.1111/2041-210X.12438}}}}
\author{Hideyasu Shimadzu\thanks{Email: \texttt{hs50@st-andrews.ac.uk}} \qquad Maria Dornelas \qquad Anne E. Magurran\\
~\\
{\it Centre for Biological Diversity and Scottish Oceans Institute},\\
{\it University of St Andrews, UK}
}
\date{}

\begin{document}






\maketitle

\renewcommand{\thefootnote}{\fnsymbol{footnote}}
\begin{abstract} \footnotetext[0]{URL: \texttt{http://synergy.st-andrews.ac.uk/diversity/}}
Range migrations in response to climate change, invasive species and the emergence of novel ecosystems highlight the importance of temporal turnover in community composition as a fundamental part of global change in the Anthropocene. Temporal turnover is usually quantified using a variety of metrics initially developed to capture spatial change. However, temporal turnover is the consequence of unidirectional community dynamics resulting from processes such as population growth, colonisation and local extinction.
Here, we develop a framework based on community dynamics, and propose a new temporal turnover measure. 
A simulation study and an analysis of an estuarine fish community both clearly demonstrate that our proposed turnover measure offers additional insights relative to spatial-context-based metrics. 
Our approach reveals whether community turnover is due to shifts in community composition or in community abundance, and identifies the species and/or environmental factors that are responsible for any change.\\
~\\
\noindent
{\bf Key words:}
Beta diversity; Biodiversity; Population dynamics; Temporal turnover.
\end{abstract}

\section{Introduction}
A large number of metrics have been developed to measure community dissimilarity \citep{Anderson2011, Jost2011, Vellend2011, Saether2013, Legendre2014a} in the century since \citet{Jaccard1901, Jaccard1912} first proposed a method of quantifying differences in assemblage composition. This variation in the identities of species found in different sites, as measured by Jaccard dissimilarity and other turnover metrics, was dubbed $\beta$-diversity by \citet{Whittaker1960, Whittaker1972}, to distinguish it from within assemblage diversity (known as $\alpha$-diversity). Although the emphasis, to date, has been on measuring $\beta$-diversity across spatial contexts, growing concern about threats to biodiversity underlines an urgent need to quantify and understand temporal turnover \citep{Dornelas2013}. Recent analyses have shown that although the $\alpha$-diversity of local communities is not changing consistently through time, with many assemblages showing increasing trends and others showing decreasing trends \citep{Vellend2013, Dornelas2014}, the rate of change in community composition (temporal $\beta$-diversity) is greater than predicted by null models of baseline temporal turnover \citep{Dornelas2014, Supp2014}. 

One approach to quantifying temporal turnover is to use metrics such as the Jaccard and Bray-Curtis \citep{Bray1957} indices, initially developed to capture spatial change. However, temporal turnover has features, such as unidirectional change, not usually present in investigations of spatial $\beta$-diversity \citep{Dornelas2013}. Moreover, temporal turnover is the consequence of community dynamics resulting from processes such as local immigration and extinction, population growth and density dependence. 
Ideally, then, temporal turnover should recognise the dimensionality of temporal change and the ecological processes that lead to shifts in community composition through time. 

Here we formalise the concept of temporal turnover. We present a novel framework for measuring temporal turnover based on the dynamics of species abundance within a community. The turnover measure we propose is directly derived from a general form of a population dynamics model which describes the extent to which each population, and consequently the community to which these species belong, changes over time. Our framework highlights the fact that the concept of turnover can be dissected into two key ecological aspects: change in community composition and change in community size (or capacity) {\it sensu} its abundance \citep{Brown1981}. In addition to providing new insights into the ecological basis of measures of temporal turnover, our approach has important practical applications because it allows the user to identify which species and/or which environmental factors play a key role in the change of the species community. We demonstrate how the turnover measure performs using a simulation and by analysing an estuarine fish community time-series from the Bristol Channel, UK. We also discuss the relationship between our approach and existing methods of $\alpha$- and $\beta$-diversities.
For $\beta$-diversity, we identify links between our measure and those previously developed to capture the nestedness and turnover components of $\beta$-diversity \citep{Baselga2010, Baselga2012, Baselga2013}.

\section{A measure of temporal turnover} \label{TOver}
The temporal changes that all ecological communities experience are a cumulative consequence of changes in the abundance of each species in the community.
Population ecology describes the dynamics of species abundance by differential equations, modelling the change rate of the population size of a given species.
We can consider this rate of change in abundance as, literally, the temporal {\it turnover} of the species in question. This assumption is the basis of our approach. The temporal turnover of the whole community can then be defined as an additive effect of turnover in each species, since the total abundance of the community is the sum of the species abundances in the community. Keeping these points in mind, we propose a new measure of temporal turnover of communities, and identify two key ecological components of this turnover: change in community composition and change in total abundance.

Throughout the paper we consider an ecological community consisting of $s$ species whose abundance varies over time governed by the population's background state at each time $t$. We use the term {\it expected abundance} to denote the background state, which is often described using other terms such as mean abundance, model expectation and (ideal) population size (a more precise definition will follow in Section \ref{estlambda}). We recognise that observed abundances
reflect both
expected abundance and natural variability. This recognition allows us, when we have observations, to separate the effect of natural variability from the expected abundance in which features of turnover can be found.
As such, here we construct a framework on the expected abundance, instead of the observations themselves.

Let $\lambda_i(t) >0$ be the expected abundance of the $i$-th species at time $t$.
Note the fact that the observed abundance can be zero even though the expected abundance is assumed to be always positive.
A species with very low expected abundance (close to zero) tends to be absent in observations most of the time.
We can therefore deal with the case of species absence, 
adopting expected abundance as the basis of our framework.
The expected total-abundance of the community is then the sum of each species' expected abundance as $\lambda(t)=\sum_{i=1}^s \lambda_i(t)$.
This fact is always valid regardless of whether the species abundances within a community are independent or not, and serves as the basis of our discussion below.

We begin by examining the dynamics of the populations that make up the community. Consider an equation for the expected abundance of the $i$-th species as
\begin{equation}
\lambda_i(t+dt) = (1+\alpha_i(t) dt) \lambda_i(t), \label{Dmodel}
\end{equation}
where $\alpha_i(t)$ is the instantaneous change rate at time $t$ that drives the increase or the decrease of the species abundance; the abundance increases  when $\alpha_i(t)>0$, and decreases when $\alpha_i(t)<0$.
If the equation \eqref{Dmodel} is solved, the Malthusian growth model \citep{Malthus1798} specifies the expected abundance, $\lambda_i(t)$, and the instantaneous change rate, $\alpha_i(t)$, becomes the growth rate of the population. The rate can also be modelled by other processes such as those, birth, death, immigration and emigration rates, $b_i(t)>0$, $d_i(t)>0$, $m_i(t)>0$ and $e_i(t)>0$ respectively, like $\alpha_i(t)=b_i(t)-d_i(t)+ m_i(t) - e_i(t)$. This of course leads $\lambda_i(t)$ to another type of model. As such, in order to keep our framework below general, we do not specify any functional form of the expected abundance, $\lambda_i(t)$, at this stage, but it will be discussed in Section \ref{estlambda}. Note that
when those rates are offset relative to each other, $\alpha_i(t)=b_i(t) - d_i(t) + m_i(t) - e_i(t)=0$, the population's abundance is stable for the time increment $dt$, since its expected abundance becomes constant, $\lambda_i(t+dt)=\lambda_i(t)$.

The reasoning given above implies that change rate is a legitimate indicator to quantify, literally, the turnover of the $i$-th species, based on its abundance, for the time increment $dt$. However, as the change rate is in most cases unknown, it needs to be determined by the trajectory of the expected abundance, $\lambda_i(t)>0$, itself as
\begin{eqnarray*}
\alpha_i(t)dt
&=& \frac{\lambda_i(t+dt) - \lambda_i(t)}{\lambda_i(t)} = \frac{d \lambda_i(t)}{\lambda_i(t)} \nonumber \label{a1}\\
&=& d\log \left(\lambda_i(t) \right). \label{a2}
\end{eqnarray*}
Here and elsewhere in the paper $\log$ refers to the natural logarithms.

We now consider the community scale, represented by the expected total-abundance of the species in the community $\lambda(t)=\sum_{i=1}^s \lambda_i(t)$, applying the same analogy used for the population scale.
Although the continuous form of the expected abundances, $\lambda_i(t), i=1, 2, \ldots , s$, is unknown, we can estimate it using observations at discrete time points. We therefore rewrite Equation \eqref{Dmodel} in a discrete form representing a community dynamics model over a relatively short time interval, between times $t$ and $t+h$ ($h>0$), within which the difference equation is still valid. By integrating the instantaneous change rate, $\alpha_i(t)$, over the time interval, the community dynamics can be described as
\begin{eqnarray*}
\lambda(t+h) 
&\doteq& \sum_{i=1}^s \left(1+\int_t^{t+h} \alpha_i(v) dv \right) \lambda_i(t)\\
&=& \sum_{i=1}^s \left(1+\int_t^{t+h} d \log(\lambda_i(v)) \right) \lambda_i(t)\\
&=& \sum_{i=1}^s \left(1+\log \left(\frac{\lambda_i(t+h)}{\lambda_i(t)} \right) \right) \lambda_i(t). \label{CDmodel}
\end{eqnarray*}
The change rate and the instantaneous change rate may therefore hold the following relationship for the short time period $h$,
\begin{equation}
\frac{\lambda(t+h)-\lambda(t)}{\lambda(t)}
\doteq
\sum_{i=1}^s \log \left(\frac{\lambda_i(t+h)}{\lambda_i(t)} \right) p_i(t),
\label{eq3}
\end{equation}
where $p_i(t)$ is the relative abundance of the $i$-th species at time $t$, that is determined as
\[
p_i(t) = \frac{\lambda_i(t)}{\lambda(t)}.
\]
The collection of the relative abundances of the community --- the relative abundance distribution $\bm{p}(t)=\{p_1(t), p_2(t), \ldots , p_s(t)\}$ --- satisfies the conditions that $p_i(t) > 0$ and $\sum_{i=1}^s p_i(t)=1$.

\begin{figure}[p]
  \centering
  \includegraphics[keepaspectratio=true, width=0.6\linewidth]{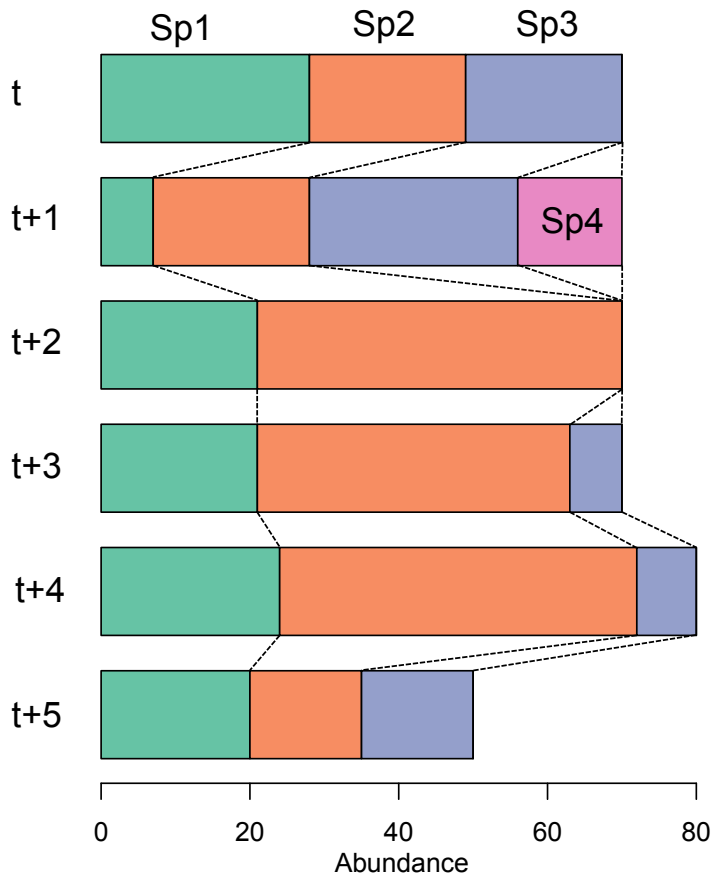}
  \caption{A schematic figure of the framework for measuring temporal turnover with four species (Sp1, Sp2, Sp3 and Sp4) in a community at times $t, t+1, \ldots , t+5$.
For clarity, the $x$-axis represents observed-abundance, but note the fact that zero observed-abundance does not necessarily mean zero expected-abundance in the framework (for details see Section \ref{estlambda}).
The proposed turnover measure $D=D_1+D_2$ (Equation \ref{eq4}) quantifies the turnover of the species community due to two key ecological processes: change in community composition ($D_1$) and change in total abundance ($D_2$).
For times $t$ to $t+3$ --- a new species Sp4 comes in at $t+1$, both Sp3 and Sp4 drop out at $t+2$ and only Sp3 comes back at $t+3$ --- the turnover is quantified by $D_1$ as there are changes in community composition but not in total abundance ($D_2=0$).
Between times $t+3$ and $t+4$, the situation is opposite: there are changes in total abundance but not in community composition. Hence, the turnover of this period is quantified by $D_2$ as $D_1=0$.
Both community composition and total abundance change between $t+4$ and $t+5$, the turnover of which is quantified by $D_1$ and $D_2$ both.
}
  \label{ExFig}
\end{figure}

Now we can formally define the turnover measure, which we call $D$. From Equation \eqref{eq3}, we define the turnover measure between times $t$ and $u, (u > t)$ as
\begin{eqnarray}
D(t : u) = \sum_{i=1}^s d_i(t : u)
&=&
\sum_{i=1}^s \log \left(\frac{\lambda_i(u)}{\lambda_i(t)}\right) p_i(t)  \label{eq3def}\\
&=&
-\sum_{i=1}^s \log \left(\frac{p_i(t)}{p_i(u)}\right) p_i(t) + \log \left(\frac{\lambda(u)}{\lambda(t)}\right)  \label{eq4}\\
&=&
D_1(\bm{p}(t) : \bm{p}(u)) + D_2(\lambda(t) : \lambda(u)) \nonumber.
\end{eqnarray}
As shown above, temporal turnover, $D$, is an abundance based measure, and is additively dissected into two quantities, $D_1$ and $D_2$.
Figure \ref{ExFig} illustrates a schematic picture of our framework for measuring the temporal turnover discussed above.

The proposed turnover measure has the following properties:
\begin{enumerate}
\item $-\infty < D < \infty$;
\item $D=0  \Longleftarrow  D_1=0$ and $D_2=0 \Longleftrightarrow  \bm{p}(t)=\bm{p}(u)$ and $\lambda(t)=\lambda(u)$;
\item $D_1 \le 0$ ({\it cf.} \citet{Kullback1951});
\item $D_2>0 \Longleftrightarrow \lambda(u)>\lambda(t)$, $D_2<0 \Longleftrightarrow \lambda(u)<\lambda(t)$.
\end{enumerate}
Note that Property 2 holds when the instantaneous change rate (Equation \ref{Dmodel}) of the every species in the community is equal to zero, $\alpha_i(t)=0, i=1, 2, \ldots , s$.

The implications of Equation \eqref{eq4} are worth consideration. First, it suggests that the temporal turnover is additively decomposed into two parts: the first term ($D_1$) related to the amount of change in community composition, and the second term ($D_2$) being dependent only on the amount of change in community size {\it sensu} its abundance.
This fact highlights two important aspects in evaluating the turnover of species community: 1) change in community composition and 2) change in total abundance.

Second, it is interesting that the Kullback--Leibler divergence \citep{Kullback1951} is subsequently derived from the definition (Equation \ref{eq3def}) as $D_1$.
Kullback--Leibler divergence quantifies the difference between two relative abundance distributions of the community at different time occasions, $\bm{p}(t)$ and $\bm{p}(u)$. Thus, the first term, $D_1$, is interpreted as the part evaluating the compositional change of the community.

The basis of our framework, equation \eqref{eq3}, highlights a key feature of temporal turnover --- the arrow of time. If $\lambda(t+h)$ and $\lambda(t)$ are swapped, the instantaneous change rate $\alpha_i(t)$ takes a different form. In other words, the interpretation of the temporal turnover measure is asymmetrical to time --- a feature that contrasts with spatial turnover. Although the proposed turnover measure, $D$, could be applied to spatial contexts, by assigning $t$ and $u$ to different locations, it is important to note that its interpretation will differ from the temporal one. This is because of the asymmetrical structure inherent in temporal turnover whereas symmetricity is usually assumed when studying spatial turnover.

\section{Specifying the expected abundance $\lambda_i(t)$} \label{estlambda}

We introduced the idea of the expected abundance earlier, but have not specified it yet.
Since we can only observe species abundances and never observe the expected abundance in ecological investigations, we need to determine the expected abundance from the observations.
Here, we denote the observed abundance of the $i$-th species at time $t$ as $n_i(t)$, lower case, and treat it as a realisation of a random variable $N_i(t)$, upper case, which follows the Poisson distribution with the time varying mean parameter $\lambda_i(t)$. We adopt the convention of using lower and upper cases for non random and random variables respectively. The expected abundance is then specified as the expectation (or the mean) of the random variable as $\lambda_i(t)=\E{N_i(t)}$.
A range of approaches are available for estimating the expected abundance, $\lambda_i(t)$, from observations, $\{n_i(t)\}$; the choice is largely dependent on the researcher and data characteristics, such as the number of observations.
We give here a short summary of common approaches, widely used in population and community ecology.

Many methods of estimating the expected abundance rely on the maximum likelihood principle.
For instance, recalling the assumption we made, an approach widely used in the ecological literature is to use the observation itself as $\hat{\lambda}_i(t) = n_i(t)$. Here, the hat sign means its estimate. Using the observation as the estimate of the expected abundance can be regarded as
the maximum likelihood estimate (MLE) when we have only one observation obtained for estimating the expected abundance at time $t$. Note that this may violate some properties of the turnover measure discussed in Section \ref{TOver}, when $n_i(t)=0$ is involved. Another choice is taking the average of the observations by assuming the expected abundance to be constant, $\lambda_i(t)=\lambda_i$, over the observation period, $t=1, 2, \ldots , t_n$, the MLE of which choice is then given as $\hat{\lambda}_i = \sum_{t=1}^{t_n} n_i(t)/t_n$.

An alternative approach is to model expected abundance, $\lambda_i(t)$, in relation to relevant environment factor(s). Although this approach requires more data than those outlined above,
a range of regression type models
are available. Examples include generalised linear models \citep[GLMs;][]{McCullagh1989} and generalised additive models \citep[GAMs;][]{Hastie1990},
that delineate the relationship between species abundance and environmental factors
as well as other species' abundances \citep{Kedem2002}.
This approach also entails specifying the link function and the distribution function that the abundance, $N_i(t)$, follows, with common examples including Poisson, Negative Binomial and others, see \citet{Zuur2009, Hilbe2011} for details. This type of modelling approach
offers more detailed insights in terms of interpreting community change as is discussed in the next section. We stress that the choice of estimation approach will depend on the extent to which information is available for study, and that $n_i(t)=0$ does therefore not necessary mean $\hat{\lambda}_i(t)=0$ unless the observation is used as the estimate.
Many other approaches are, of course, applicable and not limited to those discussed above. For example, determining the trajectory of abundance, $N_i(t)$, by (stochastic) differential equations, as did \citet{Ives2003, Mutshinda2009, Mutshinda2011}, is also a sensible approach.

\section{Identifying the influential drivers}
An advantage of
modelling
species expected abundance, $\lambda_i(t)$, is that it becomes possible to identify drivers that influence the turnover measure, $D$. Equation \eqref{eq3def} reveals how this can be done.  We illustrate this using a common model class, GLMs
with the log link function, such as Poisson and Negative Binomial distributions, as an example.

Suppose that the expected abundance of the $i$-th species, $\lambda_i(t)$, is modelled by a GLM with 
the log link
as
\begin{equation}
\log(\lambda_i(t))
=
\sum_{j=1}^m \beta_{ij} x_j(t),
\label{aGLM}
\end{equation}
where $\{x_j(t)\}$ are environmental variables, and $\{\beta_{ij}\}$ are the parameters to be estimated.
From Equations \eqref{eq3def} and \eqref{aGLM}, we have the turnover measure, $D$, described in an additive form as
\begin{equation}
D(t : u)
=\sum_{j=1}^m \sum_{i=1}^s \beta_{ij} (x_j(u) - x_j(t)) p_i(t).
\label{AddDecomp}
\end{equation}
Since the contribution of the species and the environment factors are all additive, putting
\begin{eqnarray*}
d_i(t : u)
&=&\sum_{j=1}^m \beta_{ij} (x_j(u) - x_j(t)) p_i(t)~ \rm{and}\\
d_j(t : u)
&=&\sum_{i=1}^s \beta_{ij} (x_j(u) - x_j(t)) p_i(t),
\end{eqnarray*}
the contribution ratio of the $i$-species and of the $j$-th environment variable to the turnover measure, $D$, is respectively defined as 
\begin{eqnarray}
r_i(t : u)
&=&\frac{\left|d_i(t : u)\right|}{\sum_{i=1}^s \left|d_i(t : u)\right|}, \label{cr1}\\
r_j(t : u)
&=&\frac{\left|d_j(t : u)\right|}{\sum_{j=1}^m \left|d_j(t : u)\right|}. \label{cr2}
\end{eqnarray}
These quantities show what proportion each factor contributes to the absolute amount of the turnover. Whether this type of additive decomposition is available is largely dependent on how the species expected abundance, $\lambda_i(t)$, has been modelled. 
Although our example makes use of a GLM, as long as the right-hand-side of model \eqref{aGLM} is additive in terms of the environment variables, as it is in GAMs, the additive decomposition (Equation \ref{AddDecomp}) is available, and consequently the contribution ratios, Equations \eqref{cr1} and \eqref{cr2}, can be obtained.
A key fact is that the turnover measure $D$ can be additive in terms of species $i$ {\it and} environment factor $j$, given an appropriate model form and a link function.
A similar idea can be found in calculating the contribution of individual species based on the Bray--Curtis index \citep[for example, \texttt{simper} of an R package `vegan',][]{Oksanen2014} although, calculating the contribution of environment variables may not be straightforward since the Bray--Curtis index is, in general, non additive with respect to the environment variables.

\section{Simulation study}
To illustrate how our turnover measure works, and the kind of information that can be obtained, we perform a simulation study, and compare the results with those produced using two popular turnover metrics: Jaccard \citep{Jaccard1901} and Bray--Curtis \citep{Bray1957}. We simulate an ecological community within which each species
hold inter-species relationships keeping the zero-sum community condition \citep{Hubbell2001}, apart from an imposed abrupt change
to represent a disturbance as per \citet{Dornelas2010}.

\subsection{Simulated data}
In the simulation study, we assume 
an ecological community of $30$ species ($s=30$), within which species abundances at time $t$, $\bm{N}(t) = (N_1(t), N_2(t), \ldots , N_s(t))^\top$, are dependent on each other in a sense of zero-sum community \citep{Hubbell2001}.
The abundances distribute as a multinomial distribution, $\bm{N}(t) \sim {\sf Mn}(n(t), \bm{p}(t))$, and the covariance between any pair of species, $i$ and $i^\prime$, is given as $\Cov{N_i(t), N_{i^\prime}(t)} = -n(t)p_i(t)p_{i^\prime}(t)$ for $i \neq i^\prime$.
We generate series of abundance values (200 time steps for each of the $30$ species), $\{\bm{N}(t) = \bm{n}(t): 1 \le t \le 200 \}$, for the whole community
with an artificial change taking a place at $t=100$ as
\[
\Pr \left(\bm{N}(t) = \bm{n}(t); n, \bm{p} \right) =
\begin{cases}
{\sf Mn}(n, \bm{p}) & t<100, \\
{\sf Mn}(n^\prime, \bm{p}^\prime) & t \ge 100,
\end{cases}
\]
where $n$ and $n^\prime$ are the total abundance of the community and $\bm{p}$ and $\bm{p}^\prime$ are the relative species abundances.
In other words, the simulated community is a zero-sum community whose total size is respectively $n$ and $n^\prime$ before and after the time point $t=100$.
The initial values of the relative species abundances, $\bm{p}$, are obtained using the program provided by \citet{Dornelas2010}.
We consider three types of changes in community size at time $t$: stable ($n = n^\prime$), an increase ($n < n^\prime$) and a decrease ($n > n^\prime$). For the compositional change, we consider two cases: no-alteration ($\bm{p}=\bm{p}^\prime$) and alteration ($\bm{p} \neq \bm{p}^\prime$) in the relative species abundances. The scenarios to be considered are specified by a combination of those states, but we have omitted the trivial case, stable and no-alteration ($n = n^\prime$ and $\bm{p}=\bm{p}^\prime$), so that there are five scenarios to be examined.
See \ref{simulation} for the detailed explanation.

\subsection{Calculating turnover metrics}
For each of our five simulated community time series, $\{n_i(t): 1 \le t \le 200, i=1, 2, \ldots , 30; n, n^\prime, \bm{p}, \bm{p}^\prime \}$ ($n=200, n^\prime=180, 200, 230$),
the three turnover metrics: our measure $D$, Jaccard and Bray--Curtis indices, are calculated in two different ways. One is based on a GLM and the other is based on the simulated raw data, $\hat{\lambda}_i(t)=n_i(t)$.

First, we fit a GLM with a Poisson distribution as
\begin{equation}
\log \left( \lambda_i(t) \right) = 
\theta_{i1} + \theta_{i2} I(t \ge 100),
\label{fittedmodel}
\end{equation}
where $\theta_{i1}$ and $\theta_{i2}$ are the parameters to be estimated by maximising the likelihood.
The fitted model (Equation \ref{fittedmodel}) is exact re-parametrisation of the simulation data to capture the abrupt change introduced at time $t=100$.
Once the expected abundance is estimated, $\hat{\lambda}_i(t)$, the estimated relative abundance $\hat{p}_i(t)$ can be calculated as $\hat{p}_i(t) = \hat{\lambda}_i(t)/\hat{\lambda}(t)$, where $\hat{\lambda}(t)=\sum_{i=1}^s \hat{\lambda}_i(t)$. The turnover metrics: our turnover measure, Jaccard and Bray--Curtis are then calculated. However, these indices are originally defined on observations, $\{n_i(t)\}$, so we introduce a model-based version of Jaccard ($J_{\lambda}$) and Bray--Curtis ($BC_{\lambda}$) indices for this study. See \ref{model-based}.

Second, we also calculated the three turnover metrics based on the simulated data, in other words the observation $n_i(t)$ itself, $\hat{\lambda}_i(t)=n_i(t)$. In doing so, we adopt the convention $\log(0/p_i(u))0=0$ for $D_1$ of Equation \eqref{eq4} when the $i$-th species is newly observed in the community at time $u$, and also we omit the case where the $i$-th species becomes absent at time $u$, $\log(p_i(t)/0)p_i(t)=\infty$, although this treatment may conceal the cases $D_1=-\infty$, it is still useful for illustration purposes.
Alternatively, an arbitrary very small value, $\varepsilon \approx 0$, could be used for $\hat{\lambda}_i(t)$ when $n_i(t)=0$ to avoid numerical difficulties.

\subsection{Result}
\begin{figure}[p]
  \centering
  \includegraphics[keepaspectratio=true, width=0.8\linewidth]{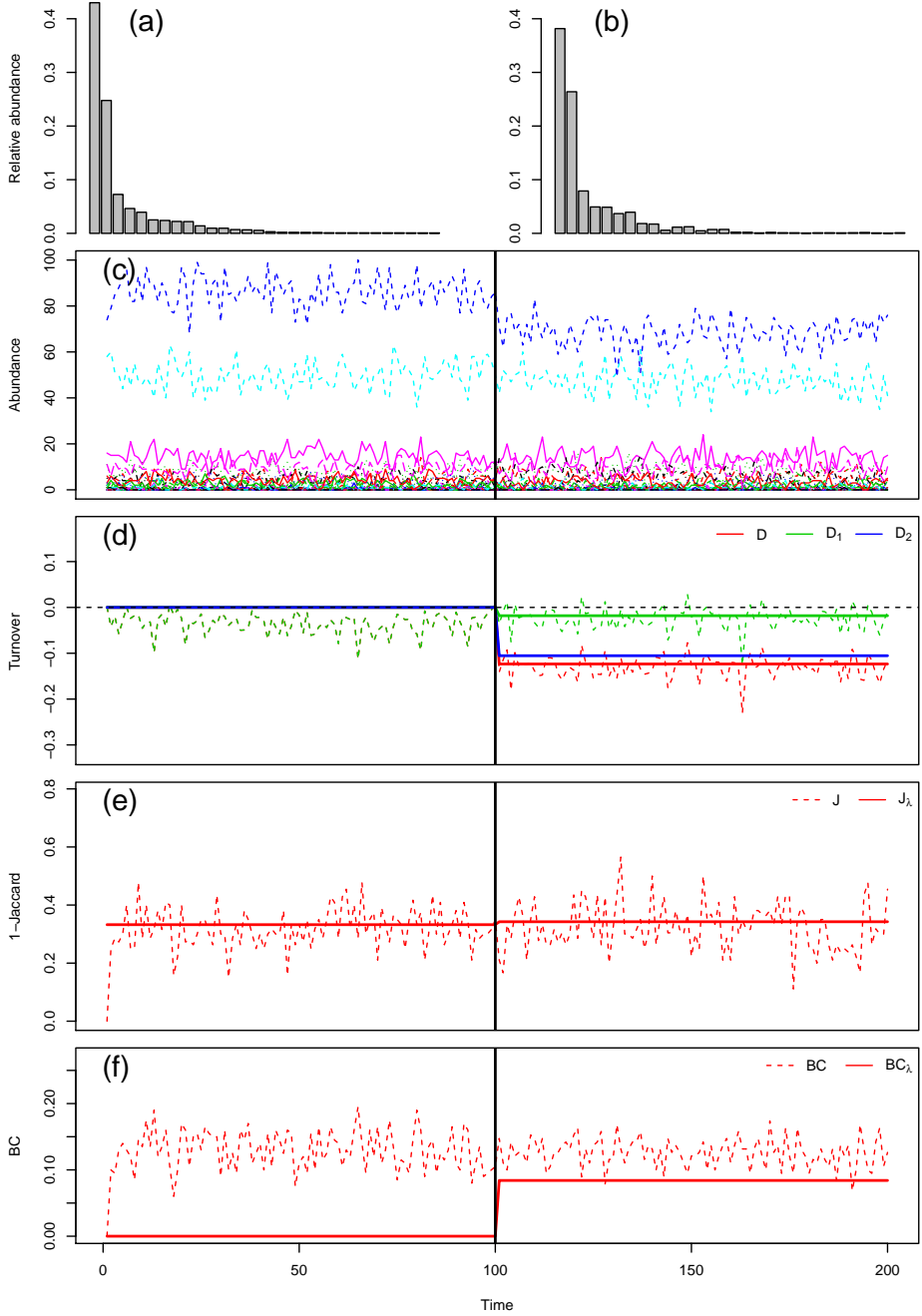}
  \caption{Simulation results of the scenario with a decrease ($n=200 > n^\prime=180$) and alteration ($\bm{p} \neq \bm{p}^\prime$). All turnover measures are calculated against the first time point. The top row shows the relative abundances before (a) and after (b) the artificial change introduced at $t=100$; (c) the simulated respective abundance series over 200 time steps; (d) the new turnover measure, $D$ (red), $D_1$ (green) representing the change in the community composition and $D_2$ (blue) representing the change in the community abundance. The solid lines are calculated based on the model (Equation \ref{fittedmodel}), and the dashed lines are based on the simulated data; (e) the outcome of Jaccard index, $1-J$ (dashed line, simulated data based) and $1-J_\lambda$ (solid line, model based); (f) the outcome of Bray--Curtis index, $BC$ (dashed line, simulated data based) and $BC_\lambda$ (solid line, model based).}
  \label{sim5}
\end{figure}

For illustration purposes, here we show the result (Figure \ref{sim5}) of the scenario,
a decrease in the total abundance ($n=200 > n^\prime=180$) and an alteration ($\bm{p} \neq \bm{p}^\prime$) in the relative species abundances.
See Figures \ref{sim1}--\ref{sim4} in \ref{simulation} for other scenarios.
Since we are interested in documenting cumulative changes, all turnover metrics are calculated against the first point of the time series (i.e. the baseline).
To reduce sensitivity to the exact nature of the baseline chosen, taking an average over some time points could also be a possible option.

In this simulation, the Jaccard index of dissimilarity, calculated on the raw simulated data ($1-J$) and on the model ($1-J_\lambda$), reveals no turnover of the community structure.
In contrast, the Bray--Curtis index, $BC$ (original metric as in \citet{Bray1957}) and $BC_\lambda$ (as in \ref{model-based}) does detect a shift in turnover. This contrast reflects the fact that species richness of the simulated community has remained relatively constant throughout, the dominant species are observed constantly but the low abundance species are observed occasionally, whereas the abundances of the species within the community shifted at the perturbation point, $t=100$. This analysis also highlights differences between the metrics based on the observations and those based on the model.
The model-based ones tend to be robust against the variation due to observations, a feature that could be useful for detecting this kind of abrupt change.

Our new turnover measure offers additional insights relative to the Jaccard and Bray--Curtis indices discussed above.
$D_1$ (the green line) quantifies the change in the community composition. With $D_1$ the absolute divergence from zero indicates change has occurred in the relative abundance distribution of the community. For example, the addition of an invasive species to a community, leading to the local extinction of a native species would be detected by this metric. The change in community abundance is indicated by $D_2$ (the blue line). When $D_2$ is positive the community abundance has increased, and when it is negative community abundance has decreased.
For those scenarios involving decreased community abundance, $n > n^\prime$ (Figures \ref{sim5} and \ref{sim1} in \ref{simulation}), $D_2$ is actually negative suggesting the community abundance decreased, whereas the Bray--Curtis index struggles to show such a difference, whether the community abundance is increased or decreased, because of it accounts for absolute differences.
Once again the model-based measure of turnover, $D$ and $D_2$, provide a clear indication of the perturbation.
A mass mortality event associated with a chemical spill, for example, would be detected by this metric.

\section{Application}
\subsection{Data}
We further explore the performance of our new turnover measure using an exceptionally complete estuarine community time series. Sampling took place at Hinkley Point `B' power station on the southern bank of the Bristol Channel in Somerset,
UK ($51^{\circ}14'14.05''$N, $3^{\circ}8'49.71''$W). Monthly quantitative sampling of a fish community commenced in January 1981, and more than 80 species have been observed over the last three decades (1981 -- 2012) \citep{Henderson1991, Henderson2005, Henderson2007, Henderson2010, Henderson2011}. Ambient environmental factors, namely water temperature and tide height, have also been recorded. See \citet{Henderson2010} for details of the survey and its methodology.

\subsection{Model}
Following \citet{Shimadzu2013}, we fit the same GAMs to the 45 core species that are consistently present in the assemblage \citep{Magurran2003}. The GAM for each $i$-th abundant species is
\[
\log(\lambda_i(t)) = \beta_{i0} + f_{i1}(\texttt{Year}) + f_{i2}(\texttt{Tide.height}) + f_{i3}(\texttt{Water.temp}) + f_{i4}(\texttt{Month}),
\]
where the $\beta_{i0}$ is a constant and $f_{ij}(\cdot)$ is a smoothing
spline function.
For the remaining 36 species that occur infrequently we fit a constant model as the simplest model,
\[
\log(\lambda_i(t)) = \beta_{i0}.
\]
Once the expected abundance is estimated, $\hat{\lambda}_i(t)$, the estimated relative abundance $\hat{p}_i(t)$ can then be calculated as $\hat{p}_i(t) = \hat{\lambda}_i(t)/\hat{\lambda}(t)$, where $\hat{\lambda}(t)=\sum_{i=1}^s \hat{\lambda}_i(t)$.

\subsection{Result}
\begin{figure}[p]
  \centering
  \includegraphics[keepaspectratio=true, width=0.8\linewidth]{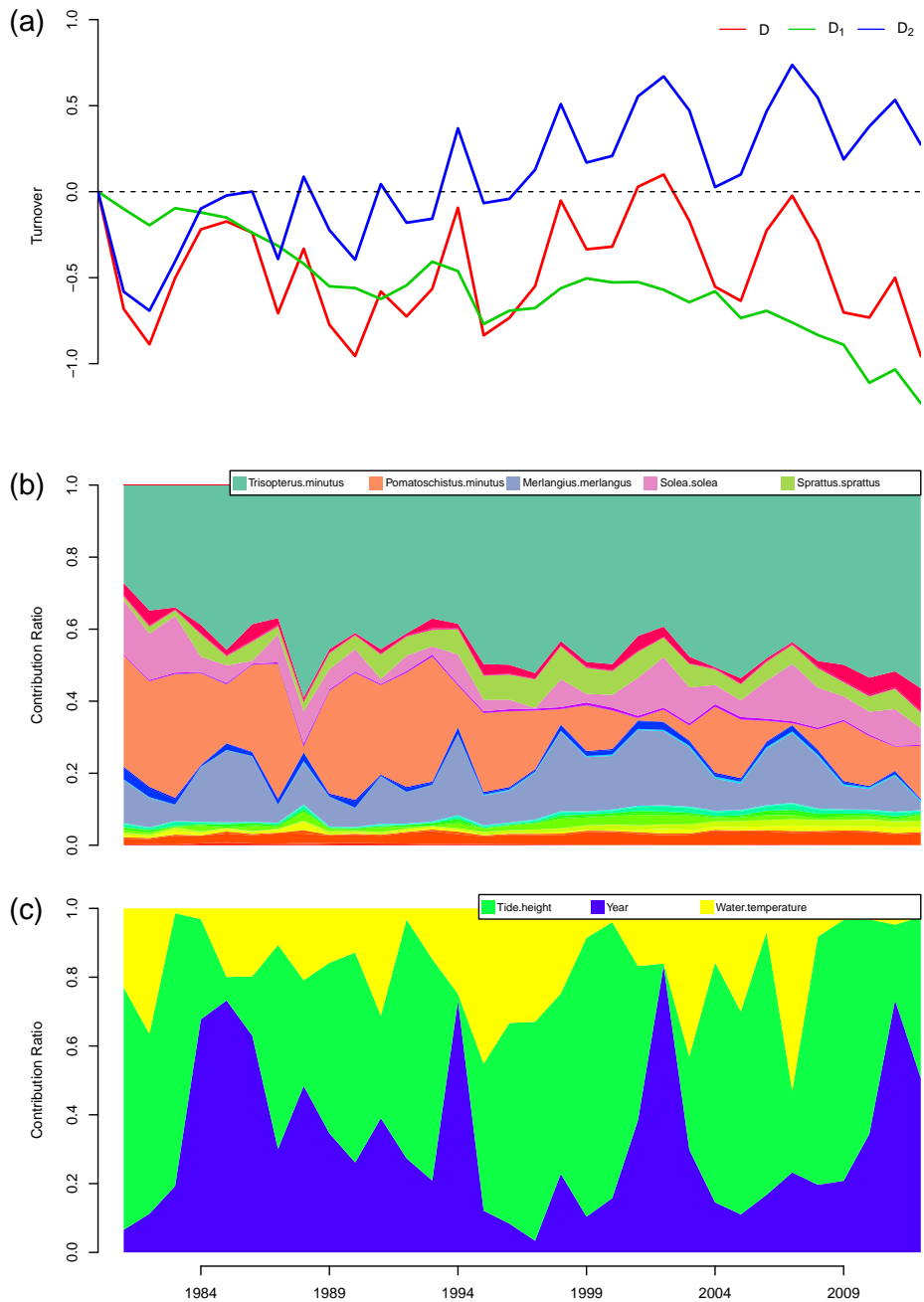}
  \caption{Illustrative results of the turnover analysis on the fish community at Hinkley Point in August over the period (1981 -- 2012). Each year is compared against the first observation, August 1981. (a) The turnover measure ($D$: the red line) and its components, the composition change ($D_1$: the green line) and the community size change ($D_2$: the blue line); (b) The contribution ratios of each species in the community. Top five species with high contribution ratio are listed in the legend, and other colour/shading represent(s) the remaining 76 species; (c) the contribution ratios of each environmental factors: year (blue), tide height (green) and water temperature (yellow).}
  \label{HinkPlot}
\end{figure}

For illustration purposes, we present the turnover result for the month August over the period (Figure \ref{HinkPlot}).
The top panel (Figure \ref{HinkPlot}a) shows the extent to which the fish community changes each year relative to the first observation, August 1981.
The turnover measure $D$ (the red line) shows a reasonably stable fluctuation a little below the zero level over the period, and also suggests cyclical variation, since the late 1990s. 
The two components of the turnover measure, one related to the change in community composition $D_1$ (the green line) and the other related to the change in community abundance $D_2$ (the blue line), tell a more detailed story. 
The green line ($D_1$) shows the changes in the relative abundance distribution of the community, quantified as the departure from zero. The continuous decrease indicates that the species composition of the community has been gradually changing since 1981. On the other hand, the blue line ($D_2$) resembles the cyclic variation of the red line ($D$) but with an increase over the period. This indicates that the community size has been getting slightly larger, relative to the 1981 level, since the late 1990s.

In Figure \ref{HinkPlot}b, the contribution ratios of some species resemble the cyclical variation in $D$ and $D_2$. The middle panel illustrates each species' contribution ratio defined by Equation \eqref{cr1}. Each coloured area represents the proportion of contribution to the change of the turnover measure $D$ made by the species.
A closer look reveals at least five species to be influential in the community: {\it Trisopterus minutus}, {\it Pomatoschistus minutus}, {\it Merlangius merlangius}, {\it Solea solea} and {\it Sprattus sprattus} (see the legend in Figure \ref{HinkPlot}b). The key role of these five species is reflected in their large contribution ratio (Equation \ref{cr1}), inducing the departure of the turnover measure $D$ from the zero level (Figure \ref{HinkPlot}a).
The large area in Figure \ref{HinkPlot}b assigned to {\it Trisopterus minutus} means that it has been particularly influential over the period. Interestingly, {\it Pomatoschistus minutus} and {\it Merlangius merlangius} appear to be offset --- as one increases in contribution ratio the other decreases, and their cyclical variations seem to correspond to the turnover measure $D$ and $D_2$ in the late 1990s. {\it Sprattus sprattus}, a known dominant species in the community, has a largely constant contribution since the late 1980s.

The bottom panel (Figure \ref{HinkPlot}c) shows the contribution ratio of the three environmental factors: year (blue), tide height (green) and water temperature (yellow). The influence of each factor varies through time.
The peaks in the year effect match with those of $D$ and $D_1$, implying the fact that this community has been under the strong influence by some particular speices whose abundance fluctuates in a relatively regular cycle.

\section{Relationship with $\alpha$-diversity measures}
The framework that we have proposed is conceptually linked with $\alpha$-biodiversity measures. Measuring biodiversity using Shannon's entropy is equivalent to examining community change rate, but between two particular occasions: the reference time and the time when the community composition has become uniform. Consider a case where the community consisting of $s$ species at the reference time $t$ and at a particular time $u, (u > t)$ when the relative abundance distribution of the community is homogeneous, $\bm{p}(u)=1/\bm{s}$. The turnover measure $D$ (Equation \ref{eq3}) is then reduced to
\[
D(t : u)
=-\log(s) - \sum_{i=1}^s p_i(t) \log (p_i(t)) + \log \left( \frac{\lambda(u)}{\lambda(t)} \right).
\]
If the community abundance stays the same, $\lambda(t)=\lambda(u)$, the turnover measure is further simplified as
\[
D(t : u)
=D_1(t : u)
=-\log(s) - \sum_{i=1}^s p_i(t) \log (p_i(t)).
\]
This is exactly Shannon's entropy with a constant shift, $-\log(s)$. In other words, the two terms on the right-hand-side are, respectively, a species richness measure and a heterogeneity measure.
Interestingly, this links back to the introduction of Shannon's entropy for measuring biodiversity by \citet{Margalef1957, Margalef1958}. Combining the above reasoning with our framework discretising differential equations, it is clear that Margalef's idea of using Shannon's entropy bridges theoretical and empirical work within biodiversity research. Theoretical population ecology has modelling of the change rate of populations as its heart. As we have explained above, empirical studies that calculate Shanon's entropy are doing something that is mathematically equivalent to investigating the change rate, but using field data.

There have been many attempts to achieve generality in the heterogeneity component of biodiversity, although by doing so the theoretical link with the community change rate as mentioned earlier is lost. \citet{Hill1973} is one of the first researchers to notice that community heterogeneity can be measured by changing the emphasis on species dominance in a community.
In terms of the turnover measure, $D$, a generalisation equivalent to \citet{Hill1973} can further be made on the change in community composition, $D_1$. Since Hill's number is the natural log scale of R\'{e}nyi's entropy \citep{Renyi1961}, R\'{e}nyi's divergence that encompasses Kullback--Leibler divergence as a special case may be used as a measure of the change in community composition, $D_1$. There have also been a variety of divergence measures examined in information theory, for example see \citet{Kawada1987, Read1988, Cichocki2010}.

$\beta$-diversity is often discussed in the context of partitioning of diversity with other types of diversity measures, such as $\gamma$- and $\alpha$-diversities \citep{Whittaker1960, Lande1996, Jost2007, Chao2012a}.
In this context \citet{Marcon2012} have derived a $\beta$-diversity measure, based on Shannon's entropy and its decomposition, using weighted Kullback--Leibler divergence. \citet{Reeve2014} have also studied diversity partitioning in the spatial context, and have discussed a $\beta$-diversity measure that considers species similarity by taking the exponent of Kullback--Leibler divergence.
However, we stress that our turnover measure, $D$, has been derived directly from the community dynamics. Moreover the derivation of $D$ is independent of biodiversity partitioning.

\section{Bray--Curtis index and its decomposition}
\begin{figure}[t]
  \centering
  \includegraphics[keepaspectratio=true, width=1.15\textwidth]{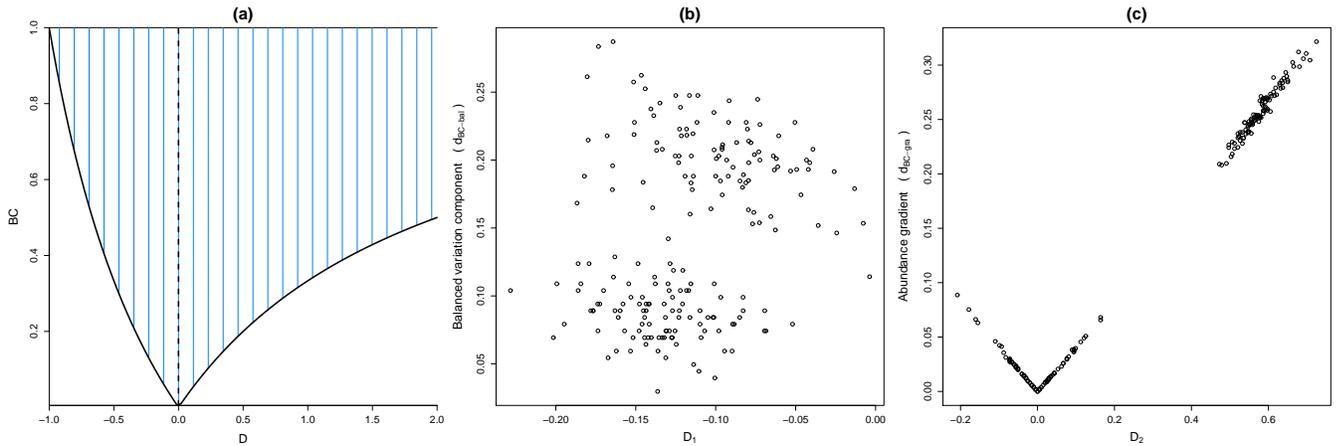}
  \caption{(a) Functional relationship of the turnover measure, $D$, and Bray--Curtis index, $BC$. The vertical striped area, $\{ (D, BC) :  \sum_{i=1}^s |d_i|/(2+D) \}$, is the values of the Bray--Curtis index can take, given the value of the turnover measure. The boundary (the black solid line) represents $BC =  |D|/(2+D)$ as $|D| \le \sum_{i=1}^s |d_i|$; (b) The relationship between the balanced variation in abundance, $d_{BC-bal}$, and the change in the species composition, $D_1$; (c) The relationship between the abundance gradient, $d_{BC-gra}$, and the the change in the community abundance, $D_2$.}
  \label{D_BC}
\end{figure}

Temporal $\beta$-diversity (temporal turnover) represents the change in biodiversity between two occasions. The Bray--Curtis index is an abundance-based metric of community dissimilarity, initially developed to measure spatial change but now increasingly used in the temporal context. Although it is, to be precise, defined using observed abundances, not on the expected abundance, there is a direct link between the turnover measure we propose and the Bray--Curtis index (see \ref{model-based} for the original definition). In fact, the Bray--Curtis index, $BC$, can be rewritten as a function of our turnover measure, $D$, as
\begin{equation}
BC \doteq \frac{\sum_{i=1}^s |d_i|}{2+D},
\label{BCD}
\end{equation}
where $|d_i|$ is the absolute value of $d_i=\log(\lambda_i(u)/\lambda_i(t))p_i(t)$, and recall that the turnover measure $D$ is the sum of these $d_i$'s over the species $i$ $(i=1, 2, \ldots , s)$ which make up the community, so that $D=\sum_{i=1}^s d_i$ (Equation \ref{eq3def}). Figure \ref{D_BC}a illustrates this functional relationship (Equation \ref{BCD}). The vertical striped area shows where the value of Bray--Curtis index ($BC$) can lie, given the value of the turnover measure ($D$), as the numerator of Equation \eqref{BCD} varies depending on the extent to which the species in the community increase or decrease. The black solid line represents the lower bound of Bray--Curtis index, the case where the abundances of all species have increased. The lower bound clearly shows that the Bray--Curtis index is asymmetrical against the community abundance change. In other words, the Bray--Curtis index tends to amplify the effect of decreases in community abundance compared to its increase.

\citet{Baselga2010, Baselga2012, Baselga2013} studied a decomposition of $\beta$-biodiversity metrics, such as Jaccard, S{\o}rensen \citep{Sorensen1948} and Bray--Curtis indices, and showed that they can be additively decomposed into two components. In particular, the Bray--Curtis index can be dissected into balanced variation in abundance and the abundance gradient components \citep{Baselga2013}. We note here that our decomposition (Equation \ref{eq4}) has parallels with but is not equivalent to these components. In fact, the balanced variation in abundance, $d_{BC-bal}$, and the abundance gradient, $d_{BC-gra}$, are functions of the turnover measure, $D=D_1+D_2$.

The balanced variation in abundance is described as
\begin{eqnarray}
d_{BC-bal}
&\doteq&
\begin{cases}
D^+ e^{-D_2}, & (D < 0)\\
~\\
-D^-, &  (D > 0)
\end{cases}
\label{bal}
\end{eqnarray}
where $D^+=\sum_{i=1}^s(\lambda_i(u)-\lambda_i(t))_+/\lambda(t)$ and $D^-=\sum_{i=1}^s(\lambda_i(u)-\lambda_i(t))_-/\lambda(t)$. Here $(\cdot)_+$ and $(\cdot)_-$ respectively denotes its positive or negative part; note that $D=D^+ + D^-$ but clearly $D_1 \neq D_2 \neq D^+$ or $D^-$. This means that the balanced variation in abundance can be interpreted as the proportion of abundance increased or decreased. Figure \ref{D_BC}b illustrates that there is little relationship between the balanced variation in abundance, $d_{BC-bal}$, and the change in the species composition, $D_1$.

The abundance gradient is also described as
\begin{eqnarray}
d_{BC-gra}
&\doteq&
\begin{cases}
\displaystyle
\frac{|D|}{2+D}qe^{-D_2}, & (D < 0)\\
~\\
\displaystyle
\frac{D}{2+D}q, &  (D > 0)
\end{cases}
\label{gra}
\end{eqnarray}
where $q=\sum_{i=1}^s \min (\lambda_i(u), \lambda_i(t))/\lambda(t)$.
This is proportional to Bray--Curtis index itself, and is its lower bound.
Figure \ref{D_BC}c illustrates the relationship between the abundance gradient, $d_{BC-gra}$, and the change in the total abundance, $D_2$.
For the detailed derivations of Equations \eqref{bal} and \eqref{gra} from \citet{Baselga2013} paper, see \ref{Baselga}.

\citet{Legendre2014} has recently reviewed approaches to partitioning $\beta$-diversity, and identified two major classes: the Baselga family that we have discussed above, and the Podani family \citep{Podani2011, Carvalho2013}; see also \citet{Legendre2013} for the discussion on partitioning $\beta$-diversity. Since the Podani family is mathematically related to the the Baselga one, it can also be described as a function of our turnover $D$. The R package `BAT' \citep{Cardoso2014} can be used to partition $\beta$-diversity.

\section{Conclusion}
We have developed a framework for measuring the temporal turnover of species communities based on community dynamics resulting from processes, such as local immigration, extinction and population growth. Our framework decomposes turnover into two classical ecological rules \citep{Brown1981}: change in the community composition associated with allocation rules and change in community abundance due to capacity rules. The new temporal turnover measure combines these two aspects in an additive manner. The formal study of the framework highlights the links with other widely used methods of measuring $\alpha$- and $\beta$-diversity. 

The performance of the new turnover measure was examined by a simulation study and the analysis of an estuarine fish community. A comparison with other common indices, namely the Jaccard and Bray--Curtis indices, demonstrated that our measure offers new insights into the ecological basis of community change. 
In addition the simulation study has clearly shown that the approach we have described in this paper can provide useful information in investigating the community change based on observational data. Furthermore, we have demonstrated that our framework offers an important practical advantage in allowing researchers to identify the species and/or environmental factors that play important roles in community change through time.

Taken together, these features of this new approach suggest that it will be a useful framework for quantifying and interpreting temporal turnover in ecological communities at a time when the natural world is facing unprecedented threats \citep{Butchart2010}.

\section*{Acknowledgements}
We are indebted to all those who have commented on previous versions of
this paper.
We acknowledge support from the European Research Council (project
BioTIME 250189) and the Royal Society. MD acknowledges funding from the
Marine Alliance for Science and Technology Scotland (MASTS). MASTS is funded by the Scottish Funding Council (grant reference HR09011) and contributing institutions. We are grateful to Peter Henderson for allowing us to use the Hinkley Point data, and to Oscar Gaggiotti for his helpful comments on the manuscript.

The authors declare that they have no conflict of interest.

\section*{Data Accessibility}
The Hinkley Point data we used are archived by Pisces Conservation, \texttt{http://consult.pisces-conservation.com/hinkdataset.html}. These data are the property of Dr Peter Henderson and Pisces Conservation Ltd. and not to be used for commercial purposes and in any reports which are produced under a commercial contract to a third party without the explicit permission of Pisces Conservation Ltd.

\bibliographystyle{dcuDOI}
\bibliography{ref.bib}

\clearpage

\renewcommand{\thepage}{A--\arabic{page}}
\setcounter{page}{1}
\renewcommand{\thefigure}{A\arabic{figure}}
\setcounter{figure}{0}

\begin{appendix}
\renewcommand\thesection{Appendix \Alph{section}}

\section{The derivation of Equations \eqref{bal} and \eqref{gra}} \label{Baselga}
Putting $\lambda_i(u)=x_{ij}$ and $\lambda_i(t)=x_{ik}$, we have eqn 1 -- 3 in \citet{Baselga2013} as
\begin{eqnarray*}
A &=& \sum_{i=1}^s \min (\lambda_i(u), \lambda_i(t))\\
B &=& \sum_{i=1}^s \lambda_i(u) - \min (\lambda_i(u), \lambda_i(t))
= \sum_{i=1}^s (\lambda_i(u) - \lambda_i(t))_+\\
C &=& \sum_{i=1}^s \lambda_i(t) - \min (\lambda_i(u), \lambda_i(t))
= -\sum_{i=1}^s (\lambda_i(u) - \lambda_i(t))_-
\end{eqnarray*}
where $(\cdot)_+$ and $(\cdot)_-$ respectively denotes its positive or negative part. The Bray--Curtis index is then described as
\begin{eqnarray*}
d_{BC} 
&=& \frac{\sum_{i=1}^s |\lambda_i(u) - \lambda_i(t)|}{\lambda_i(u) + \lambda_i(t)}
=
\frac{B+C}{2A+B+C}\\
&=&
\frac{\min(B, C)}{A + \min(B, C)} + \frac{|B-C|}{2A+B+C} \frac{A}{A + \min(B, C)}\\
&=&
d_{BC-bal} + d_{BC-gra}.
\end{eqnarray*}
The balanced variation for each case is respectively given as
\begin{description}
\item[Case I:]$\min(B, C)=B$ ($B<C \Longleftrightarrow D < 0$)
\[
d_{BC-bal}
=
\frac{B}{A + B}
=\frac{\sum_{i=1}^s (\lambda_i(u) - \lambda_i(t))_+}{\lambda(u)}
=\frac{\sum_{i=1}^s (\lambda_i(u) - \lambda_i(t))_+}{\lambda(t)}\frac{\lambda(t)}{\lambda(u)}
\doteq D^+ e^{-D_2},
\]

\item[Case II:]$\min(B, C)=C$ ($C<B \Longleftrightarrow D > 0$)
\[
d_{BC-bal}
=
\frac{C}{A + C}
=-\frac{\sum_{i=1}^s (\lambda_i(u) - \lambda_i(t))_-}{\lambda(t)}
\doteq -D^-.
\]
\end{description}
The abundance gradient for each case is respectively given as
\begin{description}
\item[Case I:]$\min(B, C)=B$ ($B<C \Longleftrightarrow D < 0$)
\begin{eqnarray*}
d_{BC-gra}
=
\frac{|B-C|}{2A+B+C} \frac{A}{A + B}
&=&\frac{|\lambda(u)-\lambda(t)|}{\lambda(u)+\lambda(t)}\frac{\sum_{i=1}^s \min (\lambda_i(u), \lambda_i(t))}{\lambda(u)}\\
&=&\frac{|\lambda(u)-\lambda(t)|}{\lambda(t)}\frac{\lambda(t)}{\lambda(u)}\frac{\sum_{i=1}^s \min (\lambda_i(u), \lambda_i(t))}{\lambda(u)+\lambda(t)}\\
&\doteq& |D| e^{-D_2} \frac{q}{2+D},
\end{eqnarray*}

\item[Case II:]$\min(B, C)=C$ ($C<B \Longleftrightarrow D > 0$)
\begin{eqnarray*}
d_{BC-gra}
=
\frac{B-C}{2A+B+C} \frac{A}{A + C}
&=&\frac{\lambda(u)-\lambda(t)}{\lambda(u)+\lambda(t)}\frac{\sum_{i=1}^s \min (\lambda_i(u), \lambda_i(t))}{\lambda(t)}\\
&\doteq& D \frac{q}{2+D}.
\end{eqnarray*}
\end{description}

\clearpage
\section{Model-based Jaccard and Bray--Curtis indices} \label{model-based}
For the simulation study, we calculate two common indices: Jaccard \citep{Jaccard1901} and Bray--Curtis \citep{Bray1957} indices for comparison.
As these indices are originally defined on observations, $\{n_i(t)\}$, we introduce a model-based version of Jaccard and Bray--Curtis indices as follow.
\begin{description}
\item[Jaccard index ($J$)]
\[
J = \frac{\sum_{i=1}^s I(n_i(t) > 0) I(n_i(u) > 0)}{\sum_{i=1}^s I(n_i(t) + n_i(u) > 0)}
\]
\item[Model-based Jaccard index ($J_\lambda$)]
\[
J_\lambda = \frac{\sum_{i=1}^s P(N_i(t) > 0) P(N_i(u) > 0)}{\sum_{i=1}^s P(N_i(t) + N_i(u) > 0)}
=
\frac{\sum_{i=1}^s \left(1-\exp(-\lambda_i(t)) \right) \left(1-\exp(-\lambda_i(u)) \right)}{s - \sum_{i=1}^s \exp(-\lambda_i(t)-\lambda_i(u))}
\]
\item[Bray--Curtis index ($BC$)]
\[
BC = \frac{\sum_{i=1}^s \left| n_i(t) - n_i(u) \right|}{n(t) + n(u)},
\]
where $n(t)=\sum_{i=1}^s n_i(t)$.
\item[Model-based Bray--Curtis index ($BC_\lambda$)]
\[
BC_\lambda = \frac{\sum_{i=1}^s \left| \lambda_i(t) - \lambda_i(u) \right|}{\lambda(t) + \lambda(u)},
\]
where $\lambda(t)=\sum_{i=1}^s \lambda_i(t)$.
\end{description}

\citet{Podani2013} note a link between Jaccard and Ruzicka \citep{Ruzicka1958} indices, taking $n_i(t)=1$ for species presence and $n_i(t)=0$ for its absence. The same link also exists between S{\o}rensen and Bray--Curtis indices, but not between those, Jaccard and Bray--Curtis indices that we have used. This type of continuity is not crucial in our study since the relative ordering is, whether Jaccard or S{\o}rensen (say $L$) index is used, retained because of their monotonic relationship, $L=2J/(J+1)$.

\clearpage
\section{Simulation results} \label{simulation}
We have studied the five simulation scenarios with the three states in changes of community size: stable ($n = n^\prime$), an increase ($n < n^\prime$) and a decrease ($n > n^\prime$); and of community composition: no-alteration ($\bm{p}=\bm{p}^\prime$) and alteration ($\bm{p} \neq \bm{p}^\prime$), when an artificial change taking a place at time $t=100$. The scenarios are specified by a combination of those states, but we have omitted the trivial case, neutral and no-alteration ($n = n^\prime$ and $\bm{p}=\bm{p}^\prime$), so that there are five scenarios to be examined. The scenario $3$ is presented in the main text.
~\\

\begin{tabular}{ccc}
\hline
Scenario & Abundance change & Composition change\\
\hline
1 & $n > n^\prime$ & $\bm{p} = \bm{p}^\prime$\\
--- & $n = n^\prime$ & $\bm{p} = \bm{p}^\prime$\\
2 & $n < n^\prime$ & $\bm{p} = \bm{p}^\prime$\\
3 & $n > n^\prime$ & $\bm{p} \neq \bm{p}^\prime$\\
4 & $n = n^\prime$ & $\bm{p} \neq \bm{p}^\prime$\\
5 & $n < n^\prime$ & $\bm{p} \neq \bm{p}^\prime$\\
\hline
\end{tabular}
~\\

For example, the interpretation of Scenario 1 here is that the total community size (abundance) decreases $n$ to $n^\prime$ by an artificial change taking a place at time $t=100$, but the community composition, the relative species abundances $\bm{p}$, stays as the same ($\bm{p} = \bm{p}^\prime$).

\begin{figure}[h]
  \centering
  \includegraphics[keepaspectratio=true, width=0.8\linewidth]{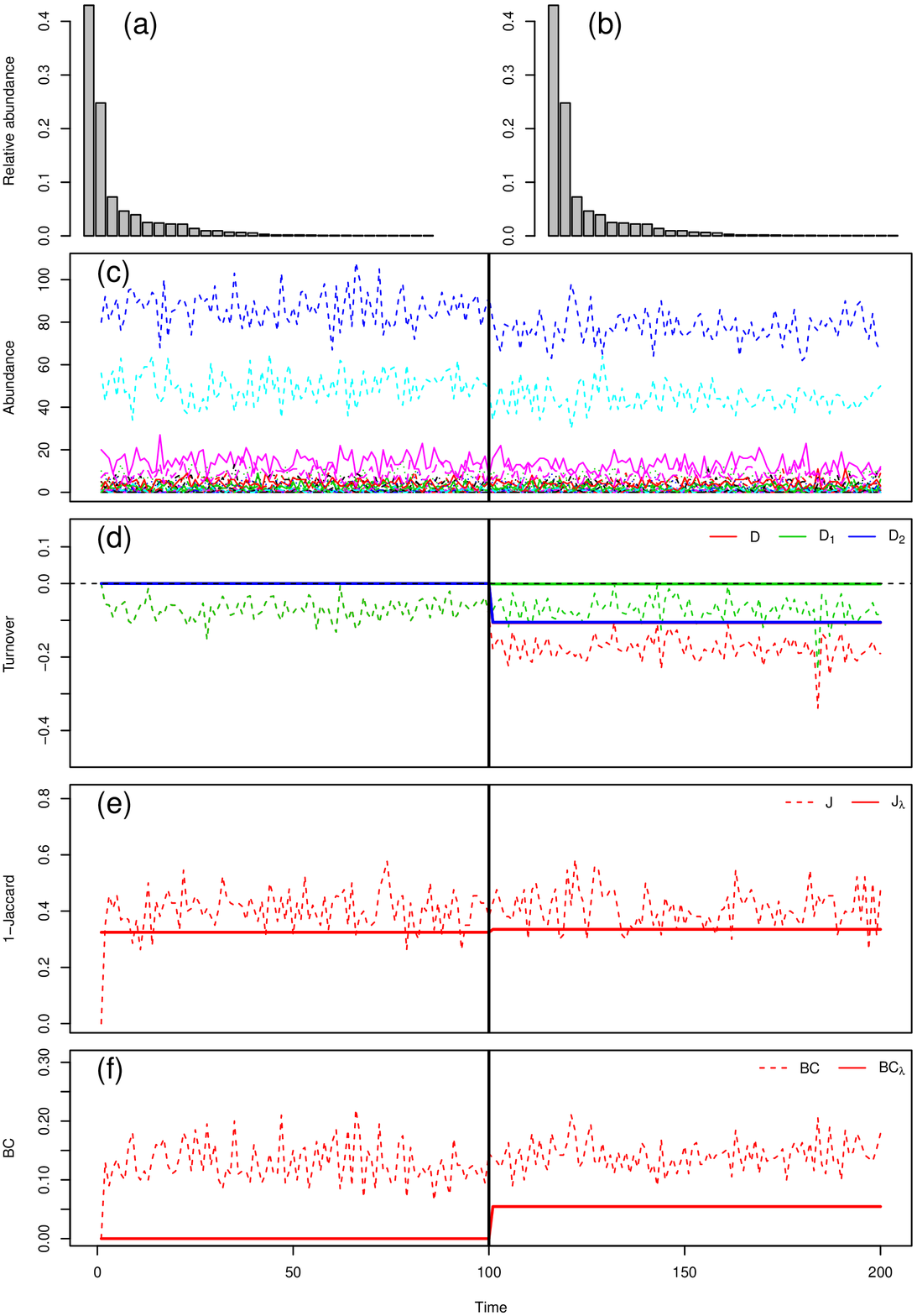}
  \caption{Simulation results of the scenario with a decrease ($n=200 > n^\prime=180$) and no-alteration ($\bm{p} = \bm{p}^\prime$). All turnover measures are calculated against the first time point. The top row shows the relative abundances before (a) and after (b) the artificial change introduced at $t=100$; (c) the simulated respective abundance series over 200 time steps; (d) the new turnover measure, $D$ (red), $D_1$ (green) representing the change in the community composition and $D_2$ (blue) representing the change in the community abundance. The solid lines are calculated based on the model (Equation \ref{fittedmodel}), and the dashed lines are based on the simulated data; (e) the outcome of Jaccard index, $1-J$ (dashed line, simulated data based) and $1-J_\lambda$ (solid line, model based); (f) the outcome of Bray--Curtis index, $BC$ (dashed line, simulated data based) and $BC_\lambda$ (solid line, model based).}
  \label{sim1}
\end{figure}

\begin{figure}[h]
  \centering
  \includegraphics[keepaspectratio=true, width=0.8\linewidth]{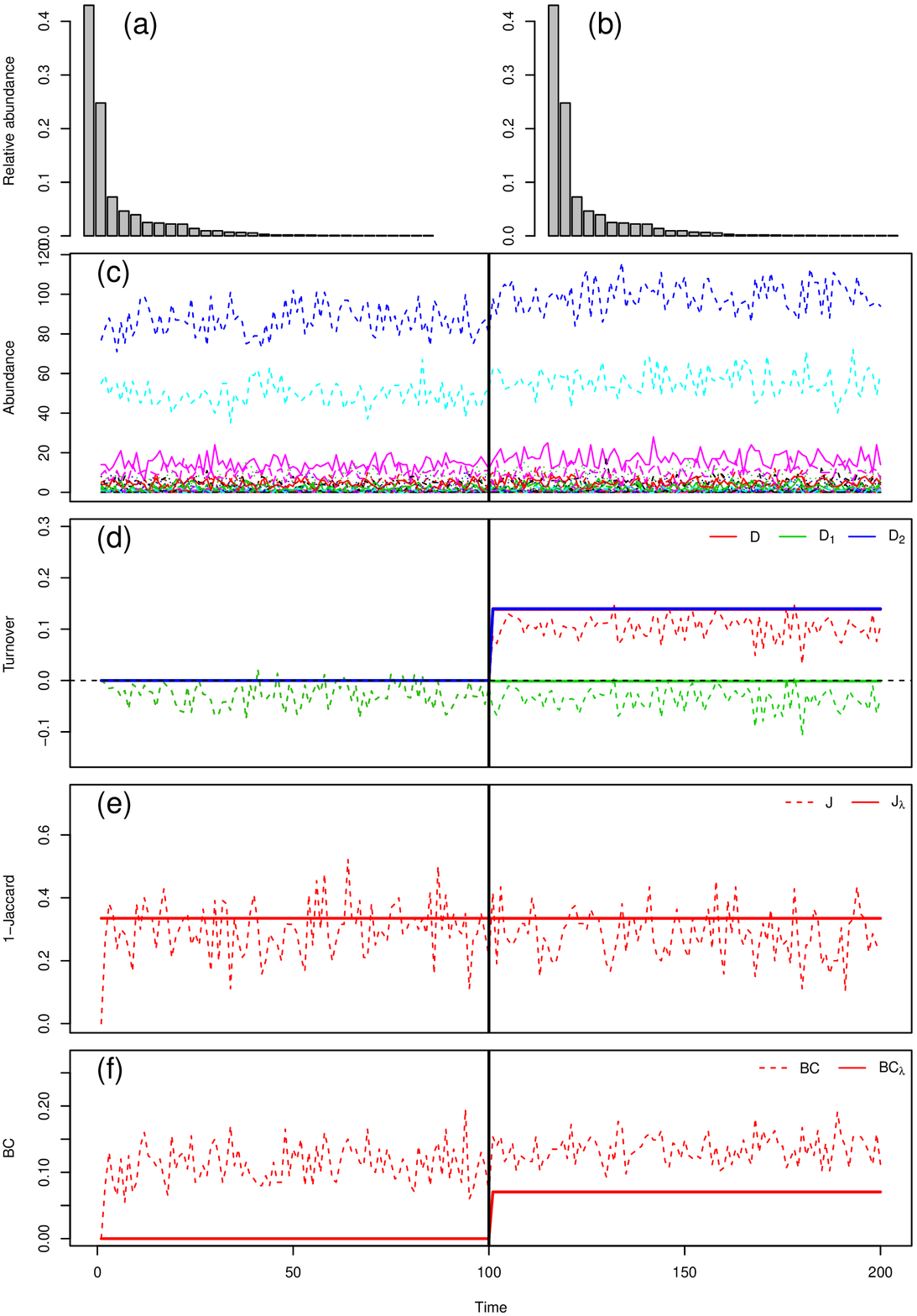}
  \caption{Simulation results of the scenario with an increase ($n=200 < n^\prime=230$) and no-alteration ($\bm{p} = \bm{p}^\prime$). All turnover measures are calculated against the first time point. The top row shows the relative abundances before (a) and after (b) the artificial change introduced at $t=100$; (c) the simulated respective abundance series over 200 time steps; (d) the new turnover measure, $D$ (red), $D_1$ (green) representing the change in the community composition and $D_2$ (blue) representing the change in the community abundance. The solid lines are calculated based on the model (Equation \ref{fittedmodel}), and the dashed lines are based on the simulated data; (e) the outcome of Jaccard index, $1-J$ (dashed line, simulated data based) and $1-J_\lambda$ (solid line, model based); (f) the outcome of Bray--Curtis index, $BC$ (dashed line, simulated data based) and $BC_\lambda$ (solid line, model based).}
  \label{sim2}
\end{figure}

\begin{figure}[h]
  \centering
  \includegraphics[keepaspectratio=true, width=0.8\linewidth]{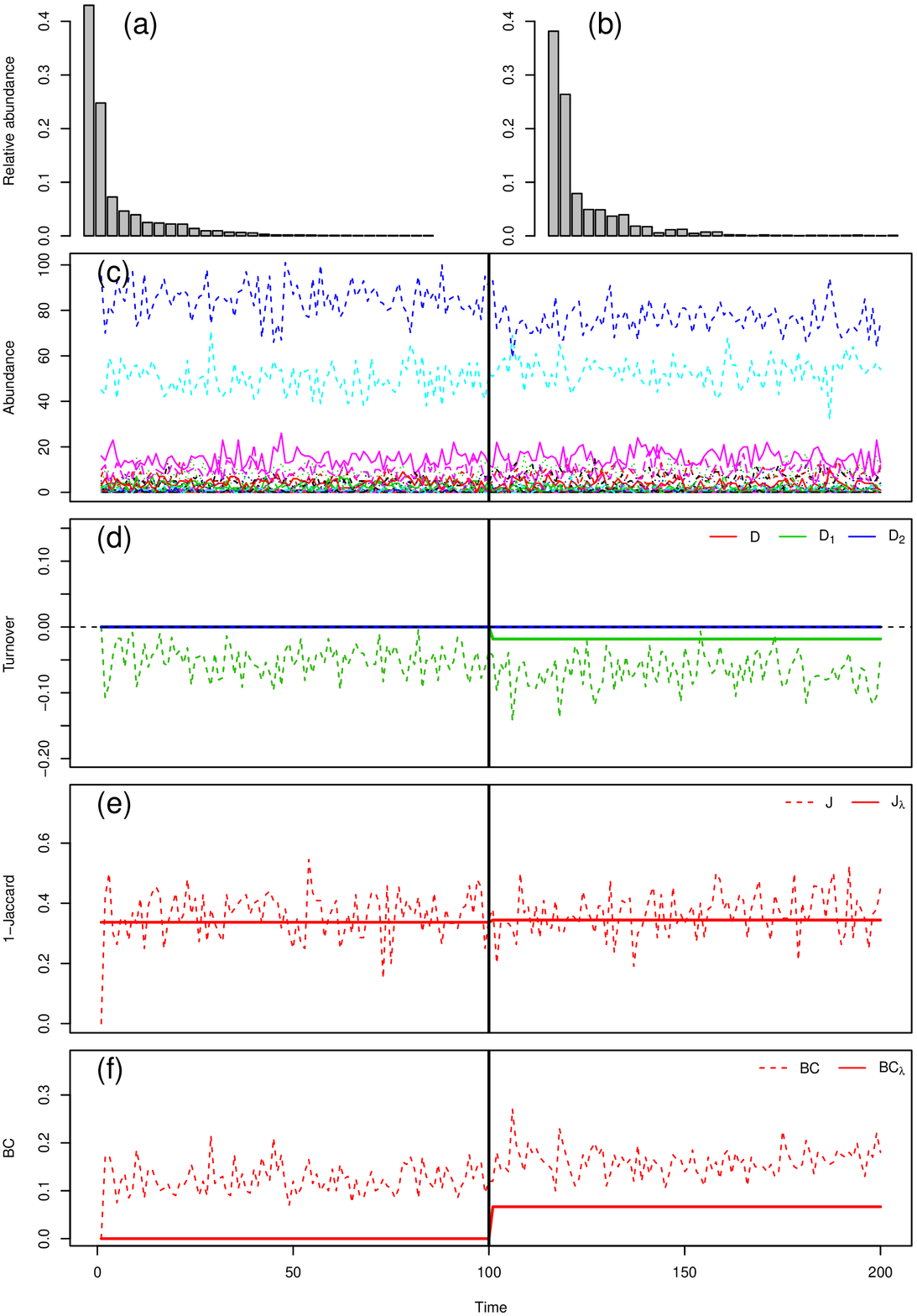}
  \caption{Simulation results of the scenario with being stable ($n=n^\prime=200$) and alteration ($\bm{p} \neq \bm{p}^\prime$). All turnover measures are calculated against the first time point. The top row shows the relative abundances before (a) and after (b) the artificial change introduced at $t=100$; (c) the simulated respective abundance series over 200 time steps; (d) the new turnover measure, $D$ (red), $D_1$ (green) representing the change in the community composition and $D_2$ (blue) representing the change in the community abundance. The solid lines are calculated based on the model (Equation \ref{fittedmodel}), and the dashed lines are based on the simulated data; (e) the outcome of Jaccard index, $1-J$ (dashed line, simulated data based) and $1-J_\lambda$ (solid line, model based); (f) the outcome of Bray--Curtis index, $BC$ (dashed line, simulated data based) and $BC_\lambda$ (solid line, model based).}
  \label{sim3}
\end{figure}

\begin{figure}[h]
  \centering
  \includegraphics[keepaspectratio=true, width=0.8\linewidth]{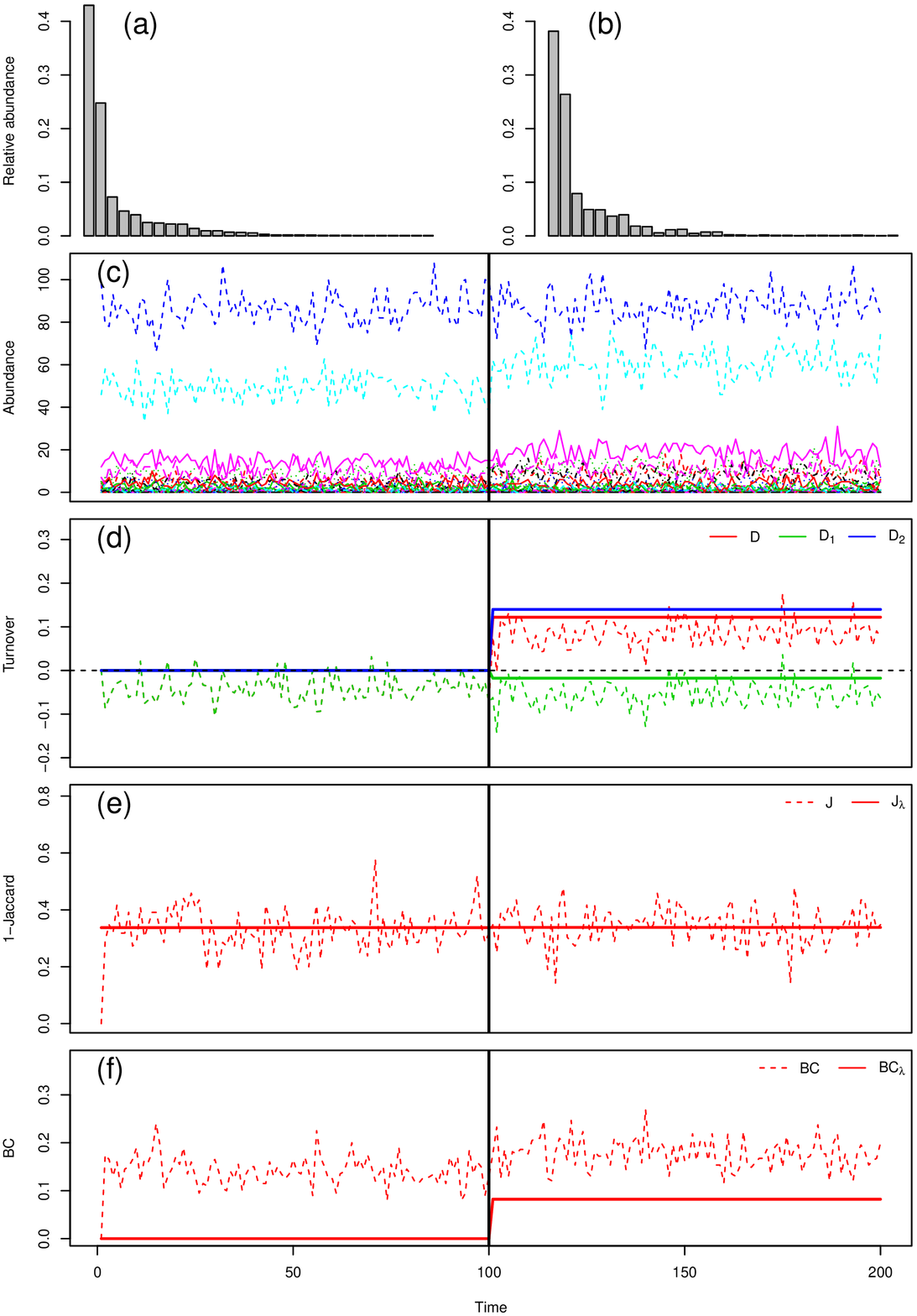}
  \caption{Simulation results of the scenario with an increase ($n=200 < n^\prime=230$) and alteration ($\bm{p} \neq \bm{p}^\prime$). All turnover measures are calculated against the first time point. The top row shows the relative abundances before (a) and after (b) the artificial change introduced at $t=100$; (c) the simulated respective abundance series over 200 time steps; (d) the new turnover measure, $D$ (red), $D_1$ (green) representing the change in the community composition and $D_2$ (blue) representing the change in the community abundance. The solid lines are calculated based on the model (Equation \ref{fittedmodel}), and the dashed lines are based on the simulated data; (e) the outcome of Jaccard index, $1-J$ (dashed line, simulated data based) and $1-J_\lambda$ (solid line, model based); (f) the outcome of Bray--Curtis index, $BC$ (dashed line, simulated data based) and $BC_\lambda$ (solid line, model based).}
  \label{sim4}
\end{figure}

\clearpage
\section{R code}
We provide R code to calculate our turnover measure and to produce the figures presented in the main text.

\subsection*{DD}
Calculates the turnover measure, $D$, $D_1$ and $D_2$ against any reference time point (the default is the first time point).
\begin{description}
\item[Usage]~\\
\verb|DD(x)|\\
\verb|DD(x, ref.t = 1, zero.rm = FALSE)|
\item[Arguments]~\\
\begin{tabular}{lp{12cm}}
\verb|x| & a data frame $(t, i)$ that contains the estimated expected abundances, $\hat{\lambda}_i(t)$.\\
\verb|ref.t| & reference time point, $t$, the turnover measure is calculated against. The default value is \verb|ref.t=1|, the first time point.\\
\verb|zero.rm| & a logical value indicating whether $p_i(u)=0$ should be stripped before the computation proceeds. The default value is \verb|zero.rm=FALSE|. To avoid any \verb|NaN| produced when observations are used as the estimated expected abundances, try \verb|zero.rm=TRUE|.
\end{tabular}
\item[Values]~\\
A data frame that contains the turnover measure $D$, $D_1$ and $D_2$ in each column.
\item[Example]~\\
\vspace{-7mm}
\begin{verbatim}
results <- list()
for(1 in 1:n) results[[i]] <- glm(y ~ x1 + x2, family="poisson")
x <- do.call(cbind, lapply(results, predict, type="response"))
DD(x)
\end{verbatim}
\item[Code]~\\
\vspace{-7mm}
\begin{verbatim}
DD <- function(x, ref.t=1, zero.rm=FALSE){
      	lmb <- apply(x, 1, sum)
      	D2 <- log(lmb/lmb[ref.t])

       x.p <- t(apply(x, 1, function(z)z/sum(z)))
       Pt <- x.p[ref.t,]
       if(zero.rm==FALSE){
         D1 <- -t(apply(x.p, 1, function(z)ifelse(Pt==0, 0, log(Pt/z)))) %*% Pt
       }else{
         D1 <- -t(apply(x.p, 1, function(z)ifelse(Pt==0|z==0, 0, log(Pt/z)))) %*% Pt
       }
              
       D <- D1 + D2
       data.frame(D, D1, D2)
      }
\end{verbatim}
\end{description}

\newpage
\subsection*{cr.cal}
Calculates the contribution ratios of the species $r_i(t, u)$ or the contribution ratios of the environment factors $r_j(t, u)$ from a list object, each branch of which must contain a GLM or a GAM object.
\begin{description}
\item[Usage]~\\
\verb|cr.cal(x)|\\
\verb|cr.cal(x, id=1, ref.t=1)|
\item[Arguments]~\\
\begin{tabular}{lp{12cm}}
\verb|x| & a list contains a GLM or a GAM output.\\
\verb|id| & a numerical value indicating whether the contribution ratios of the species, $r_i(t, u)$, (\verb|id=1|) or the contribution rations of the environmental factors, $r_j(t, u)$, (\verb|id=2|) to be calculated. The default value is \verb|id=1|.\\
\verb|ref.t| & reference time point, $t$, the turnover measure is calculated against. The default value is \verb|ref.t=1|, the first time point.
\end{tabular}
\item[Values]~\\
A data frame containing the cumulative contribution ratios that can be passed to the function \verb|cr.plot| to draw a contribution ratio diagram.
\item[Example]~\\
\vspace{-7mm}
\begin{verbatim}
x <- list()
for(1 in 1:n) x[[i]] <- glm(y ~ x1 + x2, family="poisson")
cr.cal(x)
\end{verbatim}
\item[Code]~\\
\vspace{-7mm}
\begin{verbatim}
cr.cal <- function(x, id=1, ref.t=1){
  lmb <- do.call(cbind, lapply(x, predict, type="response"))
  lmb.p <- t(apply(lmb, 1, function(z)z/sum(z)))
                                                
  lmb <- lapply(x, function(z)as.matrix(predict(z, type="terms")))
  lmb <- lapply(lmb, function(z)t(t(z)-z[ref.t,]))
  lmb <- array(unlist(lmb), dim=c(nrow(lmb[[1]]), ncol(lmb[[1]]), length(lmb)))

  if(id==1){
    absDi <- abs(apply(apply(lmb, c(1,3), sum), 1, "*", lmb.p[ref.t,])) # sp
  }else{
    absDi <- t(abs(apply(lmb, c(1,2), "%*%", lmb.p[ref.t,]))) # env
  }

  ri <- apply(absDi, 2, function(z)z/sum(z))
  ri.cum <- t(apply(ri, 2, cumsum))

  data.frame(ri.cum)
}
\end{verbatim}
\end{description}

\newpage
\subsection*{cr.plot}
Plots a contribution ratio diagram (eg. Figures \ref{HinkPlot}b and \ref{HinkPlot}c).
\begin{description}
\item[Usage]~\\
\verb|cr.plot(Y)|\\
\verb|cr.plot(Y, x=NULL, col.pal=NULL, k=5, ...)|
\item[Arguments]~\\
\begin{tabular}{lp{12cm}}
\verb|Y| & a data frame $(t, i)$ or $(t, j)$ that contains the contribution ratios of each species, $r_i(t, u)$, or each environmental factor, $r_j(t, u)$.\\
\verb|x| & the coordinates of points in the plot.\\
\verb|col.pal| & the colour palette to be used. The length needs to be the same as \verb|ncol(Y)|. The default choice is \verb|rainbow|.\\
\verb|k| & the number of species or environment factors to be listed in the legend of the figure. The default value is \verb|k=5|.\\
\verb|...| & further arguments passed to or from other methods.
\end{tabular}
\item[Example]~\\
\vspace{-7mm}
\begin{verbatim}
Y <- cr.cal(x)
names(Y) <- col.names
cr.plot(Y)
\end{verbatim}
\item[Code]~\\
\vspace{-7mm}
\begin{verbatim}
cr.plot <- function(Y, x=NULL, col.pal=NULL, k=5, ...){
  k <- min(ncol(Y), k)
  if(is.null(col.pal)){col.pal <- rainbow(ncol(Y))}
  Y <- data.frame(0, Y)
  if(is.null(x)) x <- c(1:nrow(Y))
  
  matplot(x, Y, type="n", ylab="Contribution Ratio", ...)
  for(j in 2:ncol(Y)){
    polygon(c(x, rev(x)), c(Y[,j-1], rev(Y[,j])), col=col.pal[j-1], border=NA, ...)
    }

  lgd.list <- apply(apply(Y, 1, diff, na.rm=T), 1, sum, na.rm=T)
  legend("topright", legend=names(rev(sort(lgd.list)))[1:k],
         col=col.pal[order(lgd.list, decreasing=T)][1:k],
         pch=15, pt.cex=2, bg="white", ...)
}
\end{verbatim}
\end{description}

\end{appendix}

\end{document}